\documentclass[twocolumn,preprintnumbers,amsmath,amssymb]{revtex4}

\usepackage{graphicx}
\usepackage{dcolumn}
\usepackage{bm}

\DeclareMathOperator{\erf}{erf}
\newcommand{\bra}[1]{\ensuremath{\langle #1 \vert}}
\newcommand{\ket}[1]{\ensuremath{\vert #1  \rangle}}

\renewcommand{\b}[1]{\ensuremath{\mathbf{#1}}}
\newcommand{\lr}{\ensuremath{\text{lr}}}
\newcommand{\sr}{\ensuremath{\text{sr}}}
\newcommand{\unif}{\ensuremath{\text{unif}}}
\newcommand{\Psimu}{\ensuremath{\Psi^{\lr,\mu}}}
\renewcommand{\H}{\ensuremath{\text{H}}}
\newcommand{\RPA}{\ensuremath{\text{RPA}}}
\newcommand{\pd}{\ensuremath{\text{pd}}}
\newcommand{\vir}{\ensuremath{\text{vir}}}
\renewcommand{\l}{\ensuremath{\lambda}}
\newcommand{\g}{\ensuremath{\gamma}}
\def\rv{{\bf r}}
\def\kv{{\bf k}}

\begin{document}

\title{Scaling relations, virial theorem and energy densities for long-range and short-range density functionals}

\author{Julien Toulouse}
\email{toulouse@lct.jussieu.fr}
\author{Paola Gori-Giorgi}
\email{gori@lct.jussieu.fr}
\author{Andreas Savin}
\email{savin@lct.jussieu.fr}
\affiliation{
Laboratoire de Chimie Th\'eorique, CNRS et Universit\'e Pierre et Marie Curie,\\
4 place Jussieu, 75252 Paris, France.
}

\date{\today}

\begin{abstract}
We analyze a decomposition of the Coulomb electron-electron interaction 
into a long-range and a short-range part in the framework
of density functional theory, deriving some scaling relations and the
corresponding virial theorem. We study the behavior of the local density
approximation in the high-density limit for the long-range and the short-range
functionals by carrying out a detailed analysis of the correlation energy of a uniform electron gas interacting via a long-range only electron-electron
repulsion. Possible definitions of exchange and correlation energy
densities are discussed and clarified with some examples.
\end{abstract}

\maketitle

\section{Introduction}
\label{sec:intro}

In recent years, there has been a growing interest in approaches of density functional theory (DFT)~\cite{HohKoh-PR-64} based on a long-range/short-range decomposition of the Coulomb electron-electron interaction (see, e.g., Refs.~\onlinecite{KohMeiMak-PRL-98,IikTsuYanHir-JCP-01,HeyScuErn-JCP-03,TouColSav-PRA-04,BaeNeu-PRL-05}). The idea is to use different, appropriate approximations for the long-range and the short-range contributions to the usual density functionals of the Kohn-Sham (KS) scheme~\cite{KohSha-PR-65}.

In this paper, we consider one possible long-range/short-range decomposition of the Coulombic density functionals. To gain insight into the introduced long-range and short-range density functionals, we examine some scaling relations, the virial theorem and energy densities. The local density approximation (LDA) appropriately defined for these functionals is also analyzed.

The long-range and short-range density functionals considered in this work are defined as follows (see also Refs.~\onlinecite{Sav-INC-96,LeiStoWerSav-CPL-97,PolSavLeiSto-JCP-02,PolColLeiStoWerSav-IJQC-03,SavColPol-IJQC-03,TouColSav-PRA-04,AngGerSavTou-PRA-XX}). The starting point is the decomposition of the electron-electron Coulomb interaction $w_{ee}(r) = 1/r$ as
\begin{eqnarray}
w_{ee}(r) = w_{ee}^{\lr,\mu}(r) + w_{ee}^{\sr,\mu}(r),
\end{eqnarray}
with the long-range interaction $w_{ee}^{\lr,\mu}(r) = \erf(\mu r)/r$ and the complement short-range part $w_{ee}^{\sr,\mu}(r)=1/r- \erf(\mu r)/r$. The parameter $\mu$ controls the range of the decomposition. For $\mu=0$ the long-range interaction vanishes, $w_{ee}^{\lr,\mu=0}(r)=0$, while for $\mu \to \infty$ it reduces to the Coulomb interaction, $w_{ee}^{\lr,\mu \to \infty}(r)=1/r$. The Coulombic universal density functional $F[n]$ is consequently decomposed as
\begin{eqnarray}
F[n] = F^{\lr,\mu}[n] + \bar{F}^{\sr,\mu}[n],
\end{eqnarray}
where $F^{\lr,\mu}[n]$ is a long-range universal density functional and $\bar{F}^{\sr,\mu}[n]$ is its short-range complement. The long-range functional is defined in the constrained-search formulation~\cite{Lev-PNAS-79} by
\begin{eqnarray}
 F^{\lr,\mu}[n] &=& \min_{\Psi \to n} \bra{\Psi} \hat{T}+\hat{W}_{ee}^{\lr,\mu} \ket{\Psi},
\label{Flrmin}
\end{eqnarray}
where $\hat{T}$ is the kinetic energy operator and $\hat{W}_{ee}^{\lr,\mu} = (1/2) \iint \hat{n}_2(\b{r}_1,\b{r}_2) w_{ee}^{\lr,\mu}(r_{12}) d\b{r}_1 d\b{r}_2$ the long-range interaction operator, expressed with the pair-density operator $\hat{n}_2(\b{r}_1,\b{r}_2)$.  The minimizing (multi-determinantal) wave function in Eq.~(\ref{Flrmin}) is denoted by $\Psimu[n]$. Introducing the non-interacting kinetic energy functional, $T_s[n] = \bra{\Phi[n]} \hat{T} \ket{\Phi[n]}$, where $\Phi[n]$ is the KS determinant, $F^{\lr,\mu}[n]$ is written as 
\begin{eqnarray}
 F^{\lr,\mu}[n] &=& T_s[n] + E_{\H xc}^{\lr,\mu}[n],
\label{}
\end{eqnarray}
where the long-range Hartree-exchange-correlation functional, $E_{\H xc}^{\lr,\mu}[n] = E_{\H}^{\lr,\mu}[n] + E_{x}^{\lr,\mu}[n] + E_{c}^{\lr,\mu}[n]$, is the sum of the long-range Hartree functional
\begin{eqnarray}
E_{\H}^{\lr,\mu}[n] = \frac{1}{2} \iint n(\b{r}_1) n(\b{r}_2) w_{ee}^{\lr,\mu}(r_{12}) d\b{r}_1 d\b{r}_2,
\label{EHlr}
\end{eqnarray}
the long-range exchange functional
\begin{eqnarray}
E_{x}^{\lr,\mu}[n] = \bra{\Phi[n]} \hat{W}_{ee}^{\lr,\mu} \ket{\Phi[n]} - E_{\H}^{\lr,\mu}[n],
\label{}
\end{eqnarray}
and the long-range correlation functional
\begin{eqnarray}
E_{c}^{\lr,\mu}[n] &=& \bra{\Psimu[n]} \hat{T}+\hat{W}_{ee}^{\lr,\mu} \ket{\Psimu[n]}
\nonumber\\
&&- \bra{\Phi[n]} \hat{T}+\hat{W}_{ee}^{\lr,\mu} \ket{\Phi[n]}.
\label{}
\end{eqnarray}
The short-range functional, $\bar{F}^{\sr,\mu}[n] = \bar{E}^{\sr,\mu}_{\H xc}[n] = E_{\H}^{\sr,\mu}[n] + E_{x}^{\sr,\mu}[n] + \bar{E}_{c}^{\sr,\mu}[n]$, is the sum of the short-range Hartree, exchange and correlation functionals, defined by complementarity to the Coulombic Hartree, exchange and correlation functionals, $E_{\H}[n]$, $E_{x}[n]$ and $E_{c}[n]$,
\begin{eqnarray}
E_{\H}^{\sr,\mu}[n] = E_{\H}[n] - E_{\H}^{\lr,\mu}[n],
\label{}
\end{eqnarray}
\begin{eqnarray}
E_{x}^{\sr,\mu}[n] = E_{x}[n] - E_{x}^{\lr,\mu}[n],
\label{}
\end{eqnarray}
\begin{eqnarray}
\bar{E}_{c}^{\sr,\mu}[n] = E_{c}[n] - E_{c}^{\lr,\mu}[n].
\label{}
\end{eqnarray}

The LDA approximation to the long-range exchange-correlation functional $E_{xc,{\rm LDA}}^{\lr,\mu}[n]$ is constructed from the exchange-correlation energy per electron $\varepsilon_{xc,\unif}^{\lr,\mu}$ of a uniform
electron gas interacting with potential $w_{ee}^{\lr,\mu}$
\begin{equation}
E_{xc,{\rm LDA}}^{\lr,\mu}[n]=\int n(\rv)\, \varepsilon_{xc,\unif}^{\lr,\mu}(n(\rv))\,d\rv.
\end{equation}
Similarly, the short-range LDA exchange-correlation functional $\bar{E}_{xc,{\rm LDA}}^{\sr,\mu}[n]$ is defined with the complementary quantity $\overline{\varepsilon}_{xc,\unif}^{\sr,\mu}=
\varepsilon_{xc,\unif}-\varepsilon_{xc,\unif}^{\lr,\mu}$ (see Refs.~\onlinecite{Sav-INC-96,TouSavFla-IJQC-04}).

The paper is organized as follows. In Sec.~\ref{sec:scaling}, we examine some scaling properties of the long-range and short-range functionals, and we discuss the high-density limit of the correlation functionals in LDA. In Sec.~\ref{sec:virial}, we derive the virial theorem satisfied by the long-range and short-range functionals. In Sec.~\ref{sec:epsilon}, we examine long-range and short-range energy densities. Sec.~\ref{sec:conclusion} contains our conclusions.

All the relations derived in this work are more generally true for an interaction of the form $w_{ee}^{\lr,\mu}(r) = \beta(\mu r)/r$ where $\beta$ is a continuous and differentiable function satisfying $\beta(x \to \infty) = 1$.

Atomic units (a.u.) are used throughout this work.

\section{Scaling relations}
\label{sec:scaling}

In this section, we generalize some usual scaling relations of the Coulombic density functionals~\cite{LevPer-PRA-85,Lev-PRA-91} (see also Ref.~\onlinecite{Sha-PRA-70}) to the case of the long-range and short-range density functionals. The scaled wave function of a $N$-electron system corresponding to a uniform scaling of the electron coordinates by the scale factor $\gamma >0$ is defined by (see, e.g., Refs.~\onlinecite{Foc-ZP-30,Zie-JPC-80,LevPer-PRA-85,ParYan-BOOK-89})
\begin{equation}
\label{}
\Psi_{\gamma}(\b{r}_1,...,\b{r}_N) = \gamma^{3N/2} \Psi(\gamma \b{r}_1,...,\gamma \b{r}_N).
\end{equation}
The wave function $\Psi_{\gamma}$ yields the scaled density $n_{\gamma}$
\begin{equation}
\label{eq_scaleddensity}
n_{\gamma}(\b{r})= \gamma^3 n(\gamma \b{r}).
\end{equation}

\subsection{Scaling relation for the Hartree and exchange functionals}

It can be easily verified from Eq.~(\ref{EHlr}) that the long-range Hartree functional satisfies the following scaling relation
\begin{equation}
E_{\H}^{\lr,\mu \gamma}[n_{\gamma}] = \gamma E_{\H}^{\lr,\mu}[n].
\label{}
\end{equation}
The KS determinant associated to the scaled density $n_{\gamma}$ is identical to the KS determinant associated to the density $n$ after uniform scaling of the coordinates 
\begin{equation}
\label{}
\Phi[n_{\gamma}] = \Phi_{\gamma}[n],
\end{equation}
and thus $\bra{\Phi[n_{\gamma}]} \hat{W}_{ee}^{\lr,\mu \gamma} \ket{\Phi[n_{\gamma}]} = \bra{\Phi[n]} \hat{W}_{ee}^{\lr,\mu} \ket{\Phi[n]}$, leading to the same scaling relation for the long-range exchange functional
\begin{equation}
E_{x}^{\lr,\mu \gamma}[n_{\gamma}] = \gamma E_{x}^{\lr,\mu}[n].
\label{}
\end{equation}
The short-range Hartree and exchange functionals satisfy the same scaling relations
\begin{equation}
E_{\H}^{\sr,\mu \gamma}[n_{\gamma}] = \gamma E_{\H}^{\sr,\mu}[n],
\label{}
\end{equation}
\begin{equation}
E_{x}^{\sr,\mu \gamma}[n_{\gamma}] = \gamma E_{x}^{\sr,\mu}[n].
\label{}
\end{equation}

\subsection{Scaling relation for the correlation functionals}

Let's introduce the following universal functional
\begin{eqnarray}
 F^{\lr,\mu,\l}[n] &=& \min_{\Psi \to n} \bra{\Psi} \hat{T}+ \l \hat{W}_{ee}^{\lr,\mu} \ket{\Psi},
\label{Flrlmin}
\end{eqnarray}
and the minimizing wave function is denoted by $\Psi^{\lr,\mu,\lambda}[n]$. The scaled wave function $\Psi^{\lr,\mu,\lambda}_{\gamma}[n]$ gives the density $n_{\gamma}$ and minimizes $\bra{\Psi} (1/\gamma^2) \hat{T} + (\lambda/\gamma) \hat{W}_{ee}^{\lr,\mu \gamma} \ket{\Psi}$ since $\bra{\Psi^{\lr,\mu,\lambda}_{\gamma}} (1/\gamma^2) \hat{T} + (\lambda/\gamma) \hat{W}_{ee}^{\lr,\mu \gamma} \ket{\Psi^{\lr,\mu,\lambda}_{\gamma}} = \bra{\Psi^{\lr,\mu,\lambda}} \hat{T} + \lambda \hat{W}_{ee}^{\lr,\mu} \ket{\Psi^{\lr,\mu,\lambda}}$. Consequently, doing the substitution $\lambda \to \lambda/\gamma$, the wave function $\Psi^{\lr,\mu,\lambda / \gamma}_{\gamma}[n]$ gives the density $n_{\gamma}$ and minimizes $(1/\gamma^2) \bra{\Psi} \hat{T} + \lambda \hat{W}_{ee}^{\lr,\mu \gamma} \ket{\Psi}$. Therefore, we conclude that
\begin{equation}
\Psi^{\lr,\mu\gamma,\lambda}[n_{\gamma}] = \Psi^{\lr,\mu,\lambda/\gamma}_{\gamma}[n].
\label{Psimugl}
\end{equation}
Applying this relation with $\lambda=1$, we find the following scaling relation for the long-range universal functional
\begin{equation}
F^{\lr,\mu\g}[n_\g] = \g^2 F^{\lr,\mu,1/\g}[n],
\label{Flrscaling}
\end{equation}
and consequently for the long-range correlation functional
\begin{equation}
E_c^{\lr,\mu\g}[n_\g] = \g^2 E_c^{\lr,\mu,1/\g}[n].
\label{}
\end{equation}
where $E_c^{\lr,\mu,\l}[n] = \bra{\Psi^{\lr,\mu,\lambda}[n]} \hat{T}+ \l \hat{W}_{ee}^{\lr,\mu} \ket{\Psi^{\lr,\mu,\lambda}[n]} - \bra{\Phi[n]} \hat{T}+ \l \hat{W}_{ee}^{\lr,\mu} \ket{\Phi[n]}$. The short-range correlation functional satisfies the same scaling relation
\begin{equation}
\bar{E}_c^{\sr,\mu\g}[n_\g] = \g^2 \bar{E}_c^{\sr,\mu,1/\g}[n],
\label{Ecmuscale}
\end{equation}
where $\bar{E}^{\sr,\mu,\lambda}_{c}[n] = E^{\lambda}_{c}[n] - E^{\lr,\mu,\lambda}_{c}[n]$ and $E^{\lambda}_{c}[n]$ is the Coulombic correlation functional along the linear adiabatic connection. To our knowledge, Eq.~(\ref{Psimugl}) is new while Eq.~(\ref{Flrscaling}) has already been mentioned by Yang~\cite{Yan-JCP-98}.

\subsection{High-density limit of the correlation functionals}

We study now the long-range and short-range correlation functionals under uniform scaling to the high-density limit ($\g \to \infty$). If the ground-state of the KS system is non-degenerate, $E_c^{\lr,\mu\g}[n_\g]$ goes to a constant when $\g \to \infty$
\begin{equation}
\lim_{\g \to \infty} E_c^{\lr,\mu\g}[n_\g] = \bar{E}_c^{\lr,\mu,(2)}[n],
\label{limEclrmu}
\end{equation}
where $E_c^{\lr,\mu,(2)}[n] = (1/2) ( \partial^2 E_c^{\lr,\mu,\l}[n]/\partial \l^2 )_{\l=0}$ is the second-order correlation energy in the G\"{o}rling-Levy perturbation theory~\cite{GorLev-PRA-94}, just as in the Coulombic case~\cite{Lev-PRA-91,GorLev-PRB-93}. We have a similar behavior for $\bar{E}_c^{\sr,\mu\g}[n_\g]$
\begin{equation}
\lim_{\g \to \infty} \bar{E}_c^{\sr,\mu\g}[n_\g] = \bar{E}_c^{\sr,\mu,(2)}[n],
\label{limEcsrmu}
\end{equation}
with $\bar{E}_c^{\sr,\mu,(2)}[n] = (1/2) ( \partial^2 \bar{E}_c^{\sr,\mu,\l}[n]/\partial \l^2 )_{\l=0}$.\par
It is interesting to study how the long-range and short-range LDA functionals behave
in the high-density limit, and to understand whether they can overcome some of the
well-known problems
of standard LDA in this regime~(see, e.g., \cite{PerMcMZun-PRA-81}).  
For the uniform electron gas of density $n$, the scaling transformation of Eq.~(\ref{eq_scaleddensity}) is simply written as $r_s \to r_s /\gamma$, where $r_s=(4\pi n/3)^{-1/3}$ is the Wigner-Seitz radius. The correlation energy per particle of the Coulombic electron gas diverges in the high-density limit logarithmically~\cite{GelBru-PR-57} 
\begin{equation}
\varepsilon_{c,\unif}(r_s/\g)  \mathop{\sim}_{{\g \to \infty}} -A \ln \g + \cdots,
\label{epscunifrs0}
\end{equation}
where $A=(1-\ln 2)/\pi^2$. The Coulombic LDA functional thus cannot recover the correct
high-density scaling for systems with a non-degenerate ground state~\cite{PerMcMZun-PRA-81}.  
The origin of the divergence of Eq.~(\ref{epscunifrs0})
has been widely analyzed in terms of the electron-gas pair density, both
in real and reciprocal space~(see, e.g., \cite{Kim-PRB-76,WanPer-PRB-91,PerWan2-PRB-92}), 
and, more recently,
in terms of the momentum distribution~\cite{ZieCio-PA-05}.

The investigation of the $\gamma\to\infty$ performances of the
long- and short-range LDA functionals
implies the study of the high-density behavior of the correlation energy
of an electron gas interacting with $w_{ee}^{\lr,\mu}$.
Following Refs.~\onlinecite{WanPer-PRB-91,Kim-PRB-76}, we consider
for this ``long-range'' electron gas the Fourier transform of 
the pair density, the static structure factor $S^{\lr,\mu}(k,r_s)$.
The correlation energy per particle, $\varepsilon_{c,\unif}^{\lr,\mu}$,
is given in terms of this function as
\begin{equation}
\varepsilon_{c,\unif}^{\lr,\mu}(r_s) = \frac{1}{2}\int_0^1 d\lambda
\int \frac{d\kv}{(2\pi)^3} \tilde{w}_{ee}^{\lr,\mu}(k)S_c^{\lr,\mu,\lambda}(k,r_s),
\label{eq_ecfromSc}
\end{equation} 
where $S_c^{\lr,\mu,\lambda}(k,r_s)=S^{\lr,\mu,\lambda}(k,r_s)-S_{\rm KS}(k,r_s)$ is the correlation
part of the static structure factor for the system of density parameter $r_s$ interacting with
$\lambda w_{ee}^{\lr,\mu}$, and
\begin{equation}
\tilde{w}_{ee}^{\lr,\mu}(k) = \frac{4\,\pi}{k^2}\,e^{-k^2/(4\,\mu^2)}\equiv \frac{4\,\pi}{k^2}\,\eta(k/\mu)
\label{eq_FT}
\end{equation}
is the Fourier transform of the long-range interaction $w_{ee}^{\lr,\mu}(r)=\erf(\mu r)/r$. The results
derived below hold more generally for any long-range interaction whose Fourier transform is 
of the form (\ref{eq_FT}), $\frac{4\,\pi}{k^2}\,\eta(k/\mu)$, with $\eta(x\to 0)= 1$, i.e., for any
long-range interaction whose real space form is $\beta(\mu r)/r$ with
$\beta(x\to \infty)=1$, as stated at the end of the Introduction.

\begin{figure}
\includegraphics[scale=0.70]{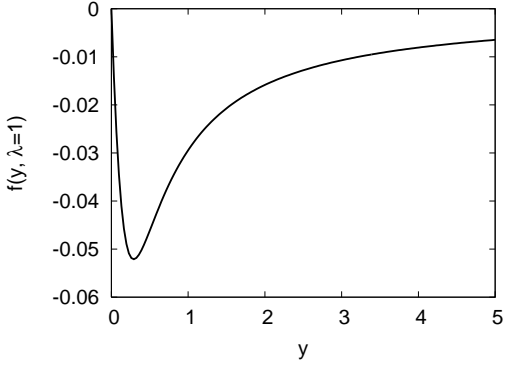}
\caption{The function $f(y,\lambda)$ of Eq.~(\ref{eq_fy}) at $\lambda=1$.}
\label{fig_fy}
\end{figure}

For the Coulomb gas,
the random-phase approximation (RPA) provides an expression
for $S_c^{\lambda}$ which is exact for small ($\ll 1$) values
of the scaled variable $q=k/2k_{\rm F}=\alpha r_s k/2$ 
(with $\alpha^3=4/9\pi$) and which gives the exact correlation energy
when $r_s\to 0$ (see, e.g., Refs.~\onlinecite{NozPin-PR-58,WanPer-PRB-91}).
The arguments for the validity of RPA 
in the high-density limit~\cite{NozPin-PR-58} can be 
extended to an interaction of the kind~(\ref{eq_FT})
as long as, when $r_s\to 0$, $\tilde{w}_{ee}^{\lr,\mu}(k)$
diverges for small $k$ as $k^{-2}$: 
in this case, in fact, the perturbation
series expansion for the correlation energy contains as leading term
an infinite number of divergent direct diagrams whose resummation yields 
the RPA expression for the correlation energy, as in the Coulomb gas. 
The RPA $S_c^{\lr,\mu,\lambda}$ reads
\begin{eqnarray}
S_{c,\RPA}^{\lr,\mu,\lambda}(q,r_s)&=&-\frac{6 q}{k_{\rm F}} \eta\left(\tfrac{2 q}{\alpha\,r_s\,\mu}\right)
\nonumber\\
&&\times
\int_0^{\infty}
\frac{\lambda\,\tilde{\chi}_0(q,iu)^2\,du}{q^2-\lambda\,\alpha\,r_s\,\tilde{\chi}_0(q,iu)\,\eta\left(\tfrac{2 q}{\alpha\,r_s\,\mu}\right)},
\nonumber\\
\label{eq_RPA}
\end{eqnarray}
where $\tilde{\chi}_0(q,iu) = (\pi/ k_{\rm F}) \chi_0(q,iu,r_s)$ is a reduced noninteracting response function, expressed in terms of the reduced imaginary frequency $u=-i\omega/(2q k_{\rm F}^2)$, that does not contain any explicit dependence on $r_s$. In the case of
the Coulombic gas (which corresponds to $\mu=\infty$, that is, $\eta=1$), 
if one simply lets $r_s$ go to zero, the  r.h.s. of Eq.~(\ref{eq_RPA}) gives
a static structure factor which behaves like $1/q$ for small $q$, thus yielding
the logarithmic divergence of the correlation energy. This divergence
comes from the combination of the two limits, $r_s\to 0$ and $q\to 0$, and requires an appropriate treatment, for which it is convenient to divide the integral of
Eq.~(\ref{eq_ecfromSc}) in two parts, $\int_0^{q_1}dq+\int_{q_1}^{\infty}dq$, with $q_1\ll 1$. The second part of this integral is finite when $r_s\to 0$, while the first part yields the logarithmic divergence and it is better studied by switching  to the scaled variable 
$y=q/\sqrt{r_s}$~\cite{Kim-PRB-76,WanPer-PRB-91,ZieCio-PA-05}, 
which gives an exact scaling for $S_c^{\lr,\mu,\lambda}$ when $q\ll 1$
and $r_s\to 0$,
\begin{eqnarray}
S_c^{\lr,\mu,\lambda}(q\ll 1,r_s\to 0) =
 \nonumber \\
 -6\sqrt{r_s}\, \alpha\, y\, \eta(\tfrac{2\,\sqrt{r_s}\, y}{\alpha\,\mu\,r_s})
\int_0^{\infty}  \frac{\lambda\,R(u)^2\, du}{y^2-\lambda\,\alpha \,R(u)\,\eta(\tfrac{2\,\sqrt{r_s}\, y}{\alpha\,\mu\,r_s})},
\label{eq_scalingRPA}
\end{eqnarray}  
where $R(u)=(u \arctan \frac{1}{u}-1)/\pi$. Equation~(\ref{eq_scalingRPA}) shows
that if we rescale $\mu$, $\mu \to \mu/r_s$, the factors $\eta(\tfrac{2\,\sqrt{r_s}\, y}{\alpha\,\mu\,r_s})$ become equal to 1 when $r_s=0$, so that the small-$q$ part of
$S_c^{\lr,\mu/r_s,\lambda}$ scales exactly to the same limit of the Coulombic gas,  i.e.,
\begin{eqnarray}
S_c^{\lr,\mu/r_s,\lambda}(q \ll 1,r_s) \mathop{\sim}_{{r_s \to 0}} -6\,\sqrt{r_s}\, \alpha\, f(y,\lambda),
\label{eq_fy}
\end{eqnarray}
where the function $f(y,\lambda)$~\cite{Kim-PRB-76,WanPer-PRB-91} behaves as $y$ for small $y$
and as $1/y$ for large $y$, and is reported in Fig.~\ref{fig_fy} in the
case $\lambda=1$.
Then everything goes as in the  Coulomb gas~\cite{WanPer-PRB-91}: in the 
small-$q$ part of the integral in Eq.~(\ref{eq_ecfromSc}) we can replace $S_c^{\lr,\mu/r_s,\lambda}$  with
Eq.~(\ref{eq_fy}), obtaining an expression of the kind $\int_0^{q_1/\sqrt{r_s}}dy f(y,\lambda)$. When
$r_s\to 0$, even if $q_1$ is small, the upper limit of this integral is large, and the $1/y$ behavior
of $f(y\gg 1)$ causes the logarithmic divergence of the correlation energy per particle as $\g \to \infty$
\begin{equation}
\varepsilon_{c,\unif}^{\lr,\mu\g}(r_s/\g)   \mathop{\sim}_{{\g \to \infty}} - A \ln \g + \cdots,
\end{equation}
with exactly the same $A$ of Eq.~(\ref{epscunifrs0}).

The short-range correlation energy per particle of the uniform electron gas
is just the difference between the correlation energies of the Coulomb
gas and of the ``long-range'' gas, so that an exact cancellation of
the logarithmic term occurs if we rescale $\mu$, 
\begin{equation}
\lim_{\g \to \infty} \bar{\varepsilon}_{c,\unif}^{\sr,\mu\g}(r_s/\g) = h(\mu, r_s).
\label{}
\end{equation}
The function $h(\mu,r_s)$
remains finite as long as $\mu r_s$ is greater than zero. This means that one can improve
the LDA performances
in the high-density limit by rescaling $\mu$ either locally,
$\mu \propto 1/r_s(\rv)$, or globally by choosing a $\mu$ bigger than $1/r_s^{\rm min}$,
where $r_s^{\rm min}$ is the minimum value of $r_s(\rv)$ in the 
given system. In other words,  the rescaled short-range functional 
allows to perform a cut-off of the correlation transferred from an electron
gas, i.e., we can eliminate the long-range correlations that occur in an electron gas but that do not occur in a confined system.

\section{Virial theorem}
\label{sec:virial}

In this section, we generalize the virial theorem of the Coulombic density functionals~\cite{LevPer-PRA-85} (see also Refs.~\onlinecite{Zie-JPC-80,AvePai-PRB-81}) to the case of the long-range and short-range density functionals.

The wave functions of the type $\Psi^{\lr,\mu}_\g[n_{1/\g}]$ give the density $n$, independently of $\g$, and can therefore be used as trial wave functions in the variational definition of $F^{\lr,\mu}[n]$ [Eq.~(\ref{Flrmin})]. As $\Psi^{\lr,\mu}_\g[n_{1/\g}]$ reduces to the minimizing wave function $\Psi^{\lr,\mu}[n]$ at $\g=1$, the stationarity condition implies that the derivative with respect to $\g$ vanishes at $\g=1$
\begin{equation}
\left( \frac{d}{d\g} \bra{\Psi^{\lr,\mu}_\g[n_{1/\g}]} \hat{T}+ \hat{W}_{ee}^{\lr,\mu} \ket{\Psi^{\lr,\mu}_\g[n_{1/\g}]} \right)_{\g=1} =0.
\label{dFdg1}
\end{equation}
The kinetic and electron-electron interaction energies have simple uniform coordinate scalings
\begin{equation}
\left( \frac{d}{d\g} (\g^2 T^{\lr,\mu}[n_{1/\g}]+ \g W_{ee}^{\lr,\mu/\g}[n_{1/\g}] )\right)_{\g=1} =0,
\end{equation}
where $T^{\lr,\mu}[n] = \bra{\Psi^{\lr,\mu}[n]} \hat{T} \ket{\Psi^{\lr,\mu}[n]}$ and $W_{ee}^{\lr,\mu}[n] = \bra{\Psi^{\lr,\mu}[n]} \hat{W}_{ee}^{\lr,\mu} \ket{\Psi^{\lr,\mu}[n]}$. Performing the derivative with respect to $\g$ leads to
\begin{eqnarray}
2 T^{\lr,\mu}[n] + W_{ee}^{\lr,\mu}[n] - \mu \frac{\partial W_{ee}^{\lr,\mu}[n]}{\partial \mu} =
\nonumber\\
 \left( \frac{d}{d\g} ( T^{\lr,\mu}[n_{\g}]+ W_{ee}^{\lr,\mu}[n_\g] )\right)_{\g=1}.
\label{TW}
\end{eqnarray}
Using the virial relation for non-interacting kinetic functional~\cite{Sha-PRA-70}
\begin{equation}
2 T_s[n] = \left( \frac{d}{d\g} T_s[n_{\g}]\right)_{\g=1},
\end{equation}
Eq.~(\ref{TW}) simplifies to
\begin{eqnarray}
T_c^{\lr,\mu}[n] + E_{\H xc}^{\lr,\mu}[n] - \mu \frac{\partial E_{\H xc}^{\lr,\mu}[n]}{\partial \mu} &=& \left( \frac{d}{d\g} E_{\H xc}^{\lr,\mu}[n_\g] \right)_{\g=1}
\nonumber\\
= -\int n(\b{r})\b{r}.\nabla \frac{\delta  E_{\H xc}^{\lr,\mu}[n]}{\delta n(\b{r})} d\b{r},
\end{eqnarray}
where $T_c^{\lr,\mu}[n] = T^{\lr,\mu}[n] - T_s[n]$ and the last equality has been obtained through a integration by parts.

It can be verified that the long-range Hartree, exchange and correlation functionals obey separate virial relations, just as for the Coulombic case,
\begin{equation}
E_{\H}^{\lr,\mu}[n] - \mu \frac{\partial E_{\H}^{\lr,\mu}[n]}{\partial \mu} = -\int n(\b{r})\b{r}.\nabla \frac{\delta  E_{\H}^{\lr,\mu}[n]}{\delta n(\b{r})} d\b{r},
\end{equation}
\begin{equation}
E_{x}^{\lr,\mu}[n] - \mu \frac{\partial E_{x}^{\lr,\mu}[n]}{\partial \mu} = -\int n(\b{r})\b{r}.\nabla \frac{\delta  E_{x}^{\lr,\mu}[n]}{\delta n(\b{r})} d\b{r},
\label{Exlrvir}
\end{equation}
\begin{equation}
T_c^{\lr,\mu}[n] + E_{c}^{\lr,\mu}[n] - \mu \frac{\partial E_{c}^{\lr,\mu}[n]}{\partial \mu} = -\int n(\b{r})\b{r}.\nabla \frac{\delta E_{c}^{\lr,\mu}[n]}{\delta n(\b{r})} d\b{r}.
\label{Eclrvir}
\end{equation}

The same relations are also valid for the short-range Hartree, exchange and correlation functionals
\begin{equation}
E_{\H}^{\sr,\mu}[n] - \mu \frac{\partial E_{\H}^{\sr,\mu}[n]}{\partial \mu} = -\int n(\b{r})\b{r}.\nabla \frac{\delta  E_{\H}^{\sr,\mu}[n]}{\delta n(\b{r})} d\b{r},
\end{equation}
\begin{equation}
E_{x}^{\sr,\mu}[n] - \mu \frac{\partial E_{x}^{\sr,\mu}[n]}{\partial \mu} = -\int n(\b{r})\b{r}.\nabla \frac{\delta  E_{x}^{\sr,\mu}[n]}{\delta n(\b{r})} d\b{r},
\label{Exsrvir}
\end{equation}
\begin{equation}
\bar{T}_c^{\sr,\mu}[n] + \bar{E}_{c}^{\sr,\mu}[n] - \mu \frac{\partial \bar{E}_{c}^{\sr,\mu}[n]}{\partial \mu} = -\int n(\b{r})\b{r}.\nabla \frac{\delta  \bar{E}_{c}^{\sr,\mu}[n]}{\delta n(\b{r})} d\b{r},
\label{Ecsrvir}
\end{equation}
where $\bar{T}_c^{\sr,\mu}[n] = T_c[n] - T_c^{\lr,\mu}[n]$ and $T_c[n]$ is the usual Coulombic correlation kinetic functional.

For the corresponding virial relations in the uniform electron gas, see Ref.~\onlinecite{Tou-PRB-XX}.

\section{Energy densities}
\label{sec:epsilon}

In this section, we examine long-range and short-range energy densities defined from pair densities or from the virial theorem. Energy densities are always useful to analyze approximations or derive new approximations (see, e.g., Refs.~\onlinecite{EngVos-PRB-93,LemRogChe-PRA-95,BurPerLev-PRA-96,BaeGri-JPCA-97,CruLamBur-JPCA-98,BurCruLam-JCP-98,ColMaySav-PRA-03}).

\subsection{Energy densities from pair densities}

In Ref.~\onlinecite{TouColSav-MP-XX}, energy densities for modified interactions defined from pair densities have been discussed. We recall that an energy density associated to the long-range exchange energy $E_x^{\lr,\mu}$ can be defined by
\begin{equation}
e^{\lr,\mu}_{x,\pd}(\b{r}) = \frac{1}{2} \int n_{2,x}(\b{r},\b{r}_{12}) w_{ee}^{\lr,\mu}(r_{12}) d\b{r}_{12},
\label{expdlr}
\end{equation}
where $n_{2,x}(\b{r},\b{r}_{12})$ is the KS exchange pair density. Likewise, the corresponding short-range energy density writes
\begin{equation}
e^{\sr,\mu}_{x,\pd}(\b{r}) = \frac{1}{2} \int n_{2,x}(\b{r},\b{r}_{12}) w_{ee}^{\sr,\mu}(r_{12}) d\b{r}_{12}.
\label{expdsr}
\end{equation}

The definition of correlation energy densities from pair densities requires an integration over an adiabatic connection. For instance, an energy density associated to the long-range correlation functional $E_c^{\lr,\mu}$ can be written as
\begin{equation}
e^{\lr,\mu}_{c,\pd}(\b{r}) = \frac{1}{2} \int_0^{\mu} d\xi \int n_{2,c}^{\lr,\xi}(\b{r},\b{r}_{12}) \frac{\partial w_{ee}^{\lr,\xi}(r_{12})}{\partial \xi} d\b{r}_{12},
\end{equation}
where $n_{2,c}^{\lr,\xi}(\b{r},\b{r}_{12})$ is the correlation pair density for the long-range interaction $w_{ee}^{\lr,\xi}(r_{12})$. The corresponding short-range correlation energy density is
\begin{equation}
\bar{e}^{\sr,\mu}_{c,\pd}(\b{r}) = \frac{1}{2} \int_{\mu}^{\infty} d\xi \int n_{2,c}^{\lr,\xi}(\b{r},\b{r}_{12}) \frac{\partial w_{ee}^{\lr,\xi}(r_{12})}{\partial \xi} d\b{r}_{12}.
\end{equation}

These energy densities involves two-electron quantities that can complicate their evaluation.

\subsection{Energy densities from the virial theorem}

Long-range and short-range energy densities can be defined from the virial theorem, just as for the Coulombic case (see, e.g., Refs.~\onlinecite{EngVos-PRB-93,CruLamBur-JPCA-98}). The virial relation of Eq.~(\ref{Exlrvir}) leads indeed to the following long-range exchange energy density
\begin{equation}
e_{x,\vir}^{\lr,\mu}(\b{r}) = - \mu \int_{\mu}^{\infty} \frac{d\xi}{\xi^2} \, n(\b{r}) \, \b{r} \cdot \nabla v_{x}^{\lr,\xi}(\b{r}),
\label{exvirlr}
\end{equation}
where $v_{x}^{\lr,\mu}(\b{r}) = \delta E_{x}^{\lr,\mu}[n]/\delta n(\b{r})$. Likewise, Eq.~(\ref{Exsrvir}) leads to the following short-range exchange energy density
\begin{equation}
e_{x,\vir}^{\sr,\mu}(\b{r}) = - \mu \int_{\mu}^{\infty} \frac{d\xi}{\xi^2} \, n(\b{r}) \, \b{r} \cdot \nabla v_{x}^{\sr,\xi}(\b{r}),
\label{exvirsr}
\end{equation}
where $v_{x}^{\sr,\mu}(\b{r}) = \delta E_{x}^{\sr,\mu}[n]/\delta n(\b{r})$. 

The virial relation of Eq.~(\ref{Eclrvir}) generalized to the linear adiabatic connection, $T_c^{\lr,\mu}[n] + E_{c}^{\lr,\mu}[n] - \mu \, \partial E_{c}^{\lr,\mu}[n]/\partial \mu = -\int n(\b{r})\b{r}.\nabla \delta E_{c}^{\lr,\mu}[n]/\delta n(\b{r}) d\b{r}$ where $T_c^{\lr,\mu}[n]$ is given by $T_{c}^{\lr,\mu,\l}[n]=E_{c}^{\lr,\mu,\l}[n] -\l \, \partial E_{c}^{\lr,\mu,\l}[n]/ \partial \l$ (see Refs.~\onlinecite{Bas-PRB-85,CruLamBur-JPCA-98}) enables to define a long-range correlation energy density
\begin{equation}
e_{c,\vir}^{\lr,\mu}(\b{r}) = - \int_{1}^{\infty} \frac{d\lambda}{\lambda^3} \, n(\b{r}) \, \b{r} \cdot \nabla v_{c}^{\lr,\mu \l,\l}(\b{r}),
\label{}
\end{equation}
where $v_{c}^{\lr,\mu \l,\l}(\b{r})=\delta E_{c}^{\lr,\mu \l,\l}[n]/ \delta n(\b{r})$. Likewise, the virial relation of Eq.~(\ref{Ecsrvir}) leads to the short-range correlation energy density
\begin{equation}
\bar{e}_{c,\vir}^{\sr,\mu}(\b{r}) = - \int_{1}^{\infty} \frac{d\lambda}{\lambda^3} \, n(\b{r}) \, \b{r} \cdot \nabla v_{c}^{\sr,\mu \l,\l}(\b{r}),
\label{}
\end{equation}
where $v_{c}^{\sr,\mu \l,\l}(\b{r})=\delta \bar{E}_{c}^{\sr,\mu \l,\l}[n]/ \delta n(\b{r})$. 

These energy densities have the advantage of involving only one-electron quantities.

\subsection{Results on the He atom}

\begin{figure}
\includegraphics[scale=0.70]{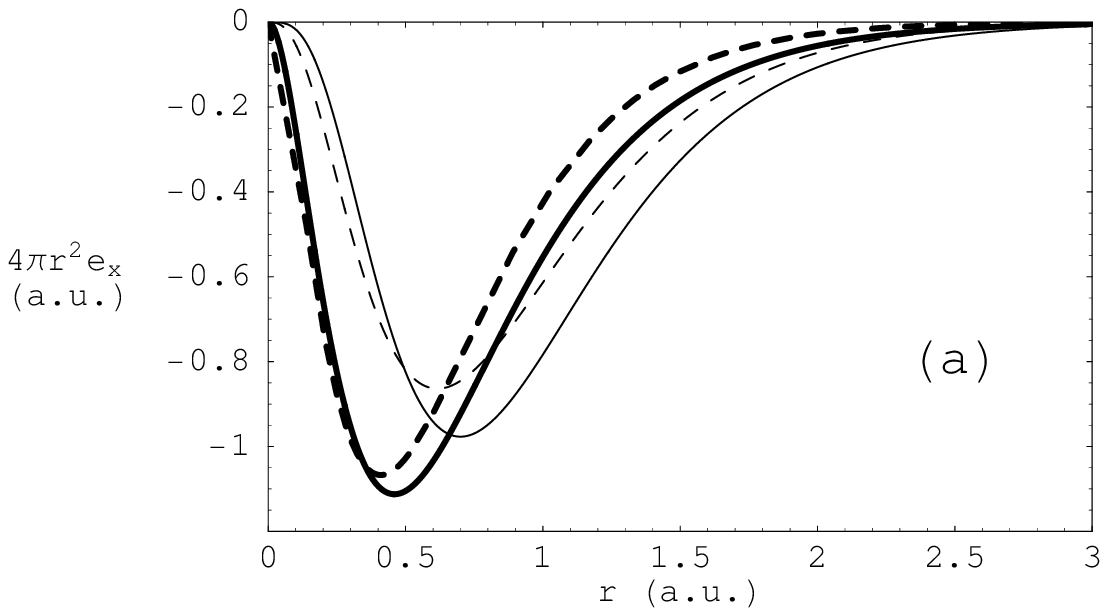}
\includegraphics[scale=0.70]{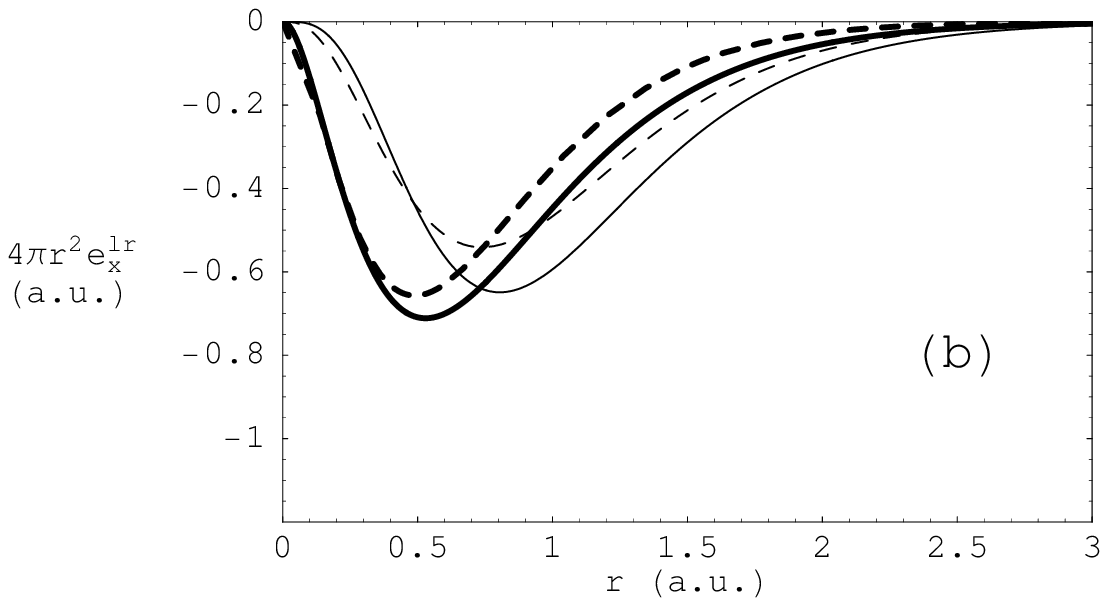}
\includegraphics[scale=0.70]{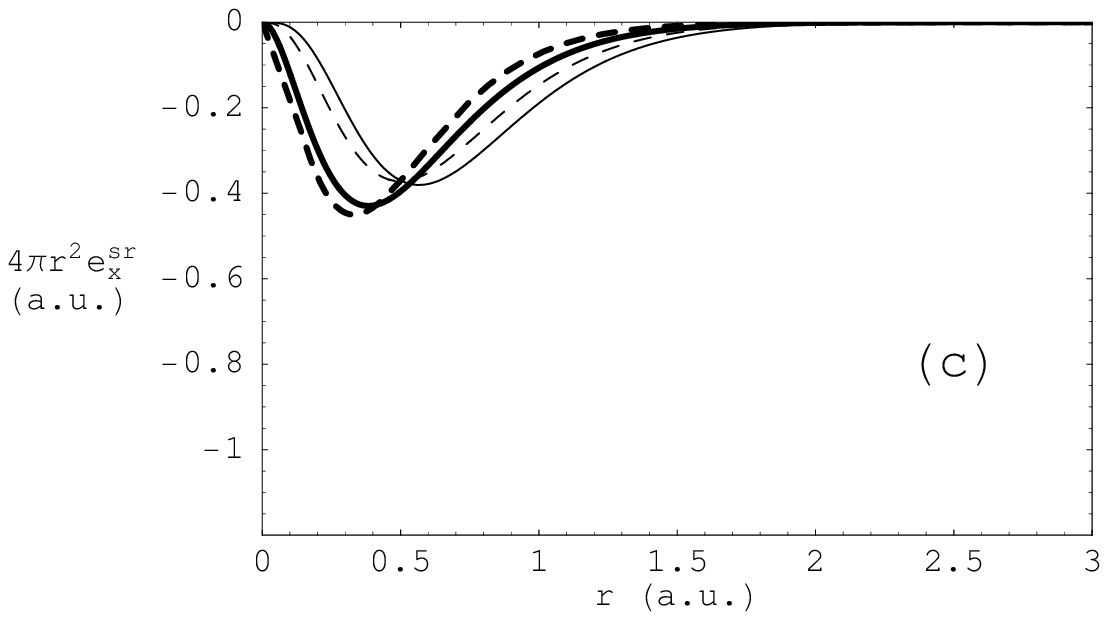}
\caption{Accurate (solid curves) and LDA (dashed curves) radial exchange energy densities defined from pair densities (thick curves) and from the virial theorem (thin curves) for the He atom: (a) Coulombic radial energy densities $4\pi r^2 e_{x,\pd}(r)$ and $4\pi r^2 e_{x,\vir}(r)$ (b) long-range radial energy densities $4\pi r^2 e_{x,\pd}^{\lr,\mu}(r)$ [Eq.~(\ref{expdlr})] and $4\pi r^2 e_{x,\vir}^{\lr,\mu}(r)$ [Eq.~(\ref{exvirlr})], (c) short-range radial energy densities $4\pi r^2 e_{x,\pd}^{\sr,\mu}(r)$ [Eq.~(\ref{expdsr})] and $4\pi r^2 e_{x,\vir}^{\sr,\mu}(r)$ [Eq.~(\ref{exvirsr})], with $\mu=1$ a.u..
}
\label{fig:expdvir}
\end{figure}

As a simple illustration, we have calculated for the He atom the long-range and short-range energy densities $e_{x,\pd}^{\lr,\mu}(r)$, $e_{x,\pd}^{\sr,\mu}(r)$, $e_{x,\vir}^{\lr,\mu}(r)$ and $e_{x,\vir}^{\sr,\mu}(r)$ for $\mu=1$ a.u., as well as the Coulombic energies densities $e_{x,\pd}(r) = e_{x,\pd}^{\lr,\mu}(r) + e_{x,\pd}^{\sr,\mu}(r)$ and $e_{x,\vir}(r) = e_{x,\vir}^{\lr,\mu}(r) + e_{x,\vir}^{\sr,\mu}(r)$. For a two-electron system, the exchange pair density and exchange potentials are directly deducible from the density. Using an accurate density, accurate exchange energy densities are thus easily obtained.

Fig.~\ref{fig:expdvir} compares the accurate and LDA radial exchange energy densities as a function of the distance to the nucleus $r$. One sees that the energy densities defined from the exchange pair density and from the virial theorem are qualitatively similar. At small $r$ ($r \lesssim 0.5$ a.u.), the LDA slightly overestimates the accurate Coulombic energy densities. At large $r$ ($r \gtrsim 0.5$ a.u.), the LDA importantly underestimates the accurate Coulombic energy densities. The contribution at large $r$ remains important in the long-range energy densities while it is significantly reduced in the short-range energy densities. Consequently, the LDA better performs for the short-range energy densities.

\begin{figure}
\includegraphics[scale=0.70]{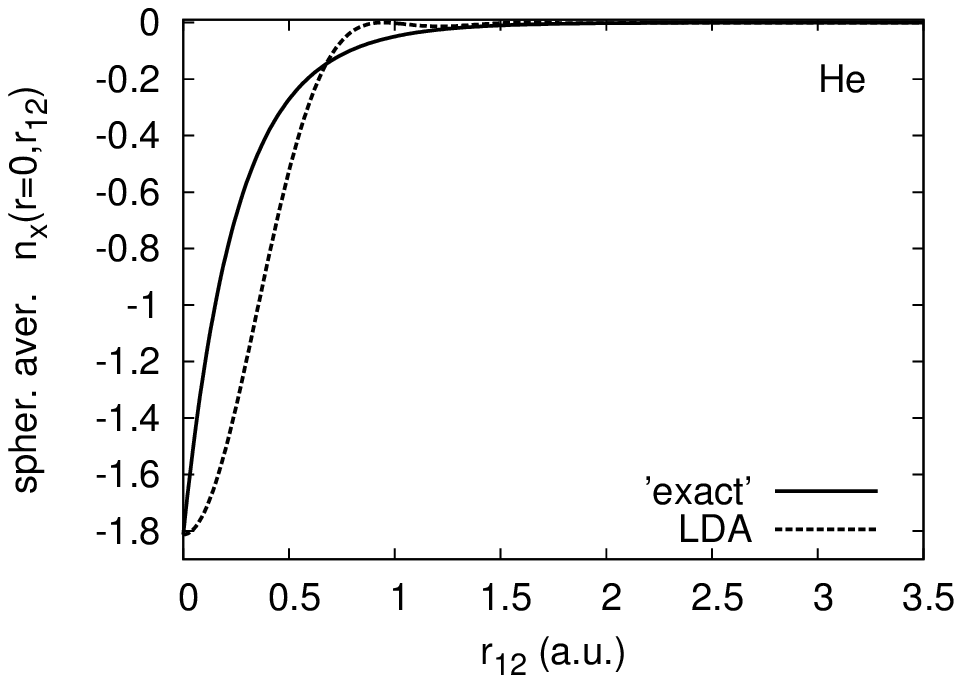}
\includegraphics[scale=0.70]{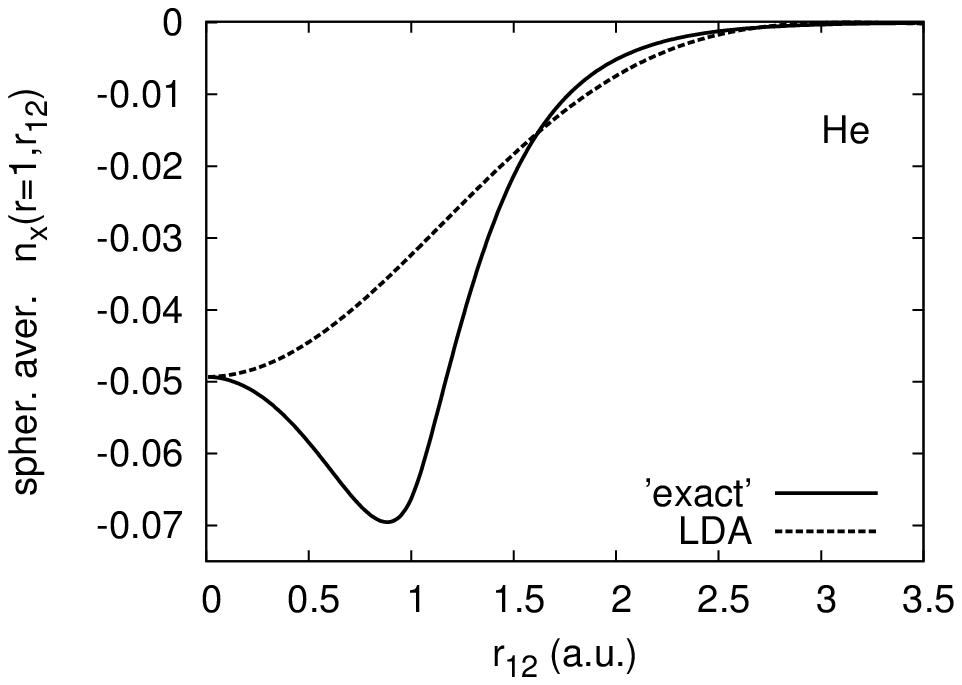}
\caption{Accurate (solid curves) and LDA (dashed curves) spherical-average exchange hole $n_{x}^{\text{sph. avr.}}(r,r_{12})$ [Eq.~(\ref{nx})] for the He atom with $r=0$ and $r=1$ a.u..
}
\label{fig:nx}
\end{figure}

In the case of the energy density defined from the exchange pair density, the better performance of the LDA at small distance $r$ can be easily explained in term of the spherical average of the exchange hole $n_{x}(r,\b{r}_{12}) = n_{2,x}(r,\b{r}_{12})/n(r)$
\begin{equation}
n_{x}^{\text{sph. avr.}}(r,r_{12}) = \frac{1}{4\pi} \int n_{x}(r,\b{r}_{12}) d\Omega_{\b{r}_{12}},
\label{nx}
\end{equation}
which is represented for the He atom in Fig.~\ref{fig:nx} with respect to the interelectronic distance $r_{12}$ for two positions of the reference electron $r=0$ and $r=1$ a.u.. For $r=0$, both the accurate and LDA exchange holes are centered at $r_{12}=0$, making the LDA a reasonable approximation. For $r=1$ a.u., the accurate hole is centered near $r_{12}=1$ a.u. while the LDA hole is still centered at $r_{12}=0$, leading to an important underestimation of the hole.

\section{Conclusions}
\label{sec:conclusion}
We have analyzed a short-range and long-range decomposition of
the Coulomb electron-electron interaction and
we have derived some exact scaling relations for the corresponding
density functionals. The study of the LDA approximation
has shown that in the high-density limit the 
short-range functional scales to a constant, thus opening
the possibility of ameliorating the performances of the
Coulomb LDA functional in this regime.
Possible definitions of energy densities obtained from pair densities and from the virial theorem have been presented. Results on the He atom suggest that the LDA approximation can give accurate short-range exchange energy densities.



\bibliographystyle{apsrev}
\bibliography{biblio}

\end{document}